\documentstyle[12pt,fleqn,epsfig]{article}

\begin{document}

\begin{titlepage}

\begin{center}
\textbf{\large Effects of orbital occupancies on the neutrinoless 
$\beta\beta$ matrix element of $^{76}$Ge}
\vspace{8mm}

\centerline{J. Suhonen$^a$, O. Civitarese$^b$}

\vspace{2mm}

\textit{ $a)$ Department of Physics, University of Jyv\"askyl\"a, 
P.O.Box 35, FIN-40014, Jyv\"askyl\"a, Finland}

\vspace{2mm}

\textit{ $b)$ Department of Physics, University of La Plata, c.c.
67, (1900) La Plata, Argentina. }

\end{center}

\vspace{4mm}
\centerline \textbf{ \today }

\vspace{4mm}

\begin{abstract}{
In this work we use the recently measured neutron occupancies in the
$^{76}$Ge and $^{76}$Se nuclei as a guideline to define the neutron 
quasiparticle states
in the 1p0f0g shell. We define the proton quasiparticles by inspecting 
the odd-mass nuclei adjacent to $^{76}$Ge and $^{76}$Se. We insert the resulting
quasiparticles in a proton-neutron quasiparticle 
random-phase approximation (pnQRPA) calculation of the
nuclear matrix element of the neutrinoless double beta ($0\nu\beta\beta$) 
decay of $^{76}$Ge. A realistic model space and effective microscopic two-nucleon 
interactions are used. We include the nucleon-nucleon short-range 
correlations and other relevant corrections at the nucleon level. It is found 
that the resulting $0\nu\beta\beta$ matrix element is smaller than in the 
previous pnQRPA calculations, and closer to the recently reported shell-model 
results.}
\end{abstract}

\vspace{0.5cm}

\noindent
PACS number(s): 21.60.Jz, 23.40.Bw, 23.40.Hc, 27.50.+e.

\vspace{0.3cm}

\noindent
Keywords: Neutrinoless double beta decay, nuclear matrix elements,
orbital occupancies

\end{titlepage}

The neutrinoless double beta ($0\nu\beta\beta$) decay of atomic nuclei
plays a key role in the search for massive Majorana neutrinos and 
their mass scale \cite{DOI85,VER86,SUH98,FAE98,KLA00,VER02}. To extract the necessary 
information from the measured data \cite{TRE02,ELL04,AVI07} the 
nuclear-structure effects have to be accounted for by computation of 
the associated nuclear matrix elements (NME's). 
Here we discuss the NME's for the light-neutrino exchange mechanism.

One most often uses the proton-neutron quasiparticle random-phase 
approximation (pnQRPA) \cite{SUH98,FAE98,TOM91} to treat the structure 
of double-beta decaying nuclei. Also recent shell-model results are 
available \cite{CAU05,CAU07,CAU08,MEN08}. 
The pnQRPA is a model tailored to describe the 
energy levels of odd-odd nuclei and their beta decays to the
neighboring even-even nuclei \cite{SUH07}. Also its extension,
renormalized pnQRPA \cite{TOI95}, has been used \cite{TOI97,ROD06} to
compute double-beta matrix elements. The pnQRPA (and the renormalized pnQRPA) 
contains a free parameter, the so-called 
particle-particle strength parameter, $g_{\rm pp}$, that
controls the magnitude of the proton-neutron two-body interactions 
in the $T=0$ pairing channel \cite{VOG86,CIV87}.
The fixing of the value of this parameter has been done either by using
the data on two-neutrino double beta ($2\nu\beta\beta$) decay
\cite{ROD06} or the data on single beta decay \cite{SUH05,CIV05}. 

The pnQRPA calculations are based on quasiparticle states \cite{SUH07} produced via
a BCS calculation \cite{SUH88}. The BCS method gives occupations of the 
single-particle orbitals and the occupation amplitudes are connected to the 
energy differences between orbitals. The customary way to determine the single-particle
energy difference is to use the Woods--Saxon mean-field potential \cite{SUH88}.
Slight adjustments of the resulting energies can be done based on the data on
energy levels of odd-mass nuclei in the neighbourhood of the nucleus where 
the pnQRPA calculation is done \cite{AUN96}. 

Recently the single-particle occupancies in the $^{76}$Ge and $^{76}$Se nuclei 
were measured by (p,t) reactions \cite{FRE07}. As a result, the vacancies 
in the neutron subspace
1p-0f$_{5/2}$-0g$_{9/2}$, hereafter called the pfg subspace, could be deduced.
In this work we exploit this spectroscopic data in pnQRPA calculations of the
nuclear wave functions of the $^{76}$As nucleus, the intermediate nucleus in
the double beta decay of $^{76}$Ge. We first produce the $2\nu\beta\beta$ NME to
fix the allowed values of the $g_{\rm pp}$ parameter by data on the half-life
of the decay \footnote{Since $^{76}$Ge lacks single beta decay
data we use the $2\nu\beta\beta$ data to fix the possible values of $g_{\rm pp}$.}. 
Based on this we finally compute the $0\nu\beta\beta$ NME's.

Here we assume that the double beta decay of $^{76}$Ge proceeds through the
virtual states of the intermediate nucleus $^{76}$As to
the ground state of the final nucleus $^{76}$Se.
By assuming the neutrino-mass mechanism to be the dominant one, we can 
write the inverse of the half-life for the $0\nu\beta\beta$ decay as \cite{SUH98}
\begin{equation} \label{eq:half}
\left[ t_{1/2}^{(0\nu)}\right]^{-1} = G_{1}^{(0\nu)}
\left( \frac{\langle m_{\nu}\rangle }{m_{\rm e}} \right)^{2}
\left( M^{(0\nu)}\right)^{2}\, ,
\end{equation}
where $m_{\rm e}$ is the electron mass and $G_{1}^{(0\nu)}$
is the leptonic phase-space factor. The $0\nu\beta\beta$
nuclear matrix element $M^{(0\nu)}$ consists of the
Gamow--Teller, Fermi and tensor parts as
\begin{equation} \label{eq:me}
M^{(0\nu)} = M_{\rm GT}^{(0\nu)} - \left( \frac{g_{\rm V}}{g_{\rm A}}
\right)^{2} M_{\rm F}^{(0\nu)} + M_{\rm T}^{(0\nu)} \ .
\end{equation}

Our numerical calculations verify the shell-model results of \cite{MEN08} and
show that the tensor part in (\ref{eq:me}) 
is quite small and its contribution can be safely neglected in what follows.
The expressions for the phase-space factor, the effective neutrino 
mass $\langle m_{\nu}\rangle$ and the matrix elements of (\ref{eq:me}) 
are given, e.g., in \cite{DOI85,SUH98,TOM91}.

\begin{table}[htb]
\caption{WS single-particle energies for the pfg subspace in units
of MeV.}
\label{t:energies}
\begin{center}\begin{tabular}{ccccc}
\hline
 & \multicolumn{2}{c}{$^{76}$Ge} & \multicolumn{2}{c}{$^{76}$Se} \\
 Orbital & neutrons & protons & neutrons & protons \\
\hline
$1\textrm{p}_{3/2}$ & $-11.52$ & $-9.015$ & $-12.77$ & $-7.000$ \\
$0\textrm{f}_{5/2}$ & $-10.72$ & $-8.212$ & $-11.98$ & $-6.288$ \\
$1\textrm{p}_{1/2}$ & $-9.797$ & $-6.789$ & $-10.98$ & $-5.001$ \\
$0\textrm{g}_{9/2}$ & $-7.030$ & $-5.311$ & $-8.305$ & $-3.371$ \\
\hline
\end{tabular}\end{center}\end{table}

The nuclear-structure calculations are performed as described in 
\cite{KOR07a,KOR07b,KOR08}.
As a model space we have used the $N=3$ and $N=4$ oscillator shells and the
0h$_{11/2}$ single-particle orbital, both for protons and neutrons. The 
starting values of the single-particle energies are obtained
from the Coulomb-corrected Woods--Saxon potential, hereafter called WS, 
with the para\-metr\-iz\-a\-tion 
of \cite{BOH69}. These energies are presented for the pfg subspace 
in Table~\ref{t:energies}. The measured neutron vacancies \cite{FRE07}
in this sub-space have been summarized in Table~\ref{t:nvac}. As can be seen from
this table, the computed vacancies in the WS basis are far from 
the measured ones: the 0g$_{9/2}$ orbital is much too thinly occupied and the 
other orbitals are too full. 

The pnQRPA calculation of the wave functions of $^{76}$As needs as input the 
neutron and proton occupation amplitudes in the pfg subspace. Instead of using
the WS based BCS amplitudes for neutrons we now resort to the occupation
amplitudes derived from the measured vacancies of Table~\ref{t:nvac}. It is 
then interesting to see to what extent this choice of occupation amplitudes 
affects the magnitudes of the computed double beta decay NME's.

We have to stress here that in this work we compute the occupancies of the 
neutron orbitals at the BCS level. Though effects beyond pairing may be present 
in the observed occupancies, in the present calculation we identify them with 
pair correlations, treated by the use of BCS. This can be justified partly by the 
good agreement between the calculated and observed properties of the low-lying
energy levels in the adjacent odd-mass nuclei. Further support can be sought
from calculations that determine the orbital occupancies and the pnQRPA 
amplitudes $X$ and $Y$ self-consistently \cite{MAR00}. In \cite{MAR00} this 
kind of scheme was applied in a realistic model space to study Fermi $pn$ 
excitations in $^{76}$Ge. From Fig. 5 of \cite{MAR00} one can see that the 
occupancies of the neutron orbitals in the pfg subspace do not change 
significantly when the BCS+pnQRPA results are replaced by the self-consistent 
results. The proton Fermi surface sharpens a bit when going from the BCS+pnQRPA 
over to the self-consistent approach of \cite{MAR00}, but the effect is 
quite small (anyway, we do not have experimental occupancies available for protons).

\begin{table}[htb]
\caption{Measured and WS based neutron vacancies in the pfg subspace.}
\label{t:nvac}
\begin{center}\begin{tabular}{ccccc}
\hline
 & \multicolumn{2}{c}{$^{76}$Ge} & \multicolumn{2}{c}{$^{76}$Se} \\
 Orbital & Exp. & WS & Exp. & WS \\
\hline
$\nu1\textrm{p}_{1/2}+\nu1\textrm{p}_{3/2}$ & 1.13 & 0.357 & 1.59 & 0.495 \\
$\nu0\textrm{f}_{5/2}$ & 1.44 & 0.500 & 2.17 & 0.618 \\
$\nu0\textrm{g}_{9/2}$ & 3.52 & 5.43 & 4.20 & 7.06 \\
\hline
\end{tabular}\end{center}\end{table}

No data on proton spectroscopic factors in the pfg subspace exist for $^{76}$Ge
or $^{76}$Se. The computed proton vacancies in the WS
basis are given in Table~\ref{t:pvac}. One possible way to access the needed
occupations for protons is to compute the spectra of the proton-odd nuclei $^{77}$As 
and $^{77}$Br, adjacent to $^{76}$Ge and $^{76}$Se. For this one can use the
quasiparticle-phonon coupling in diagonalizing the residual nuclear Hamiltonian.
If one takes the phonons to be pnQRPA phonons one ends up with a scheme called
the proton-neutron microscopic quasiparticle-phonon model (pnMQPM) \cite{MUS07}.
The pnMQPM wave function consists of a linear combination of one-quasiparticle and
three-quasiparticle components. A one-quasiparticle state in this scheme is a 
state whose wave function is dominated by a definite one-quasiparticle component.

\begin{table}[htb]
\caption{Predicted proton vacancies in the pfg subspace. The 
calculations were done in the WS and adjusted basis.}
\label{t:pvac}
\begin{center}\begin{tabular}{ccccc}
\hline
 & \multicolumn{2}{c}{$^{76}$Ge} & \multicolumn{2}{c}{$^{76}$Se} \\
 Orbital & Adj. & WS & Adj. & WS \\
\hline
$1\textrm{p}_{1/2}$ & 1.82 & 1.80 & 1.70 & 1.62 \\
$1\textrm{p}_{3/2}$ & 2.19 & 1.92 & 1.75 & 1.44 \\
$0\textrm{f}_{5/2}$ & 4.58 & 4.43 & 3.80 & 3.32 \\
$0\textrm{g}_{9/2}$ & 9.28 & 9.73 & 8.61 & 9.49 \\
\hline
\end{tabular}\end{center}\end{table}

Comparing the energies of the computed one-quasiparticle states 
with the available data gives information on the quality of the underlying BCS 
calculation and thus on the mean-field single-particle energies. The pnMQPM
computed proton one-quasiparticle states in $^{77}$As and $^{77}$Br are given
in Table~\ref{t:odda}. The wave functions corresponding to these states are indeed
dominated by a single one-quasiparticle component. This is visible from the
contribution of this component to the total normalization of the wave function,
as presented by the percentage in the last column of the table.

The predicted one-quasiparticle energies of Table~\ref{t:odda} correspond
nicely to their measured counterparts. The critical proton orbit for the success
of the pnMQPM calculation is $\pi0\textrm{g}_{9/2}$. The energy of this 
orbit had to be lowered to $-7.1\,\textrm{MeV}$ for $^{76}$Ge and
to $-5.3\,\textrm{MeV}$ for $^{76}$Se to produce a reasonable 
one-quasiparticle spectrum. The corresponding energies we call adjusted energies
and the computed proton vacancies in this basis are given in 
Table~\ref{t:pvac} in the columns called 'Adj.'. Comparison of tables
\ref{t:nvac} and \ref{t:pvac} indicates that the adjusted
proton and experimental neutron vacancies behave qualitatively
in the same way. However, the adjustments on the proton side have not such drastic
effects on the vacancies as is the case on the neutron side when going from the
WS to the experimental vacancies. For completeness, we also give
in Table~\ref{t:odda} the computed and measured energies of the neutron 
one-quasiparticle states in $^{77}$Ge and $^{77}$Se. Here the basis, adjusted 
independently of the pnMQPM calculation, shifts the $9/2^+$ state slightly
below the measured $1/2^-$ state. Related to this one has to bear in mind that
such very fine details of odd-mass spectra are hard to reproduce within 
a quasiparticle-phonon coupling scheme.

\begin{table}[htb]
\caption{One-quasiparticle states in odd-$A$ nuclei near $^{76}$Ge
and $^{76}$Se. In the last column the contribution to the total normalization 
is given in per cents.}
\label{t:odda}
\begin{center}\begin{tabular}{ccccc}
\hline
 Nucleus & State & E(exp.)[MeV] & E(th.)[MeV] & Main component(\%) \\
\hline
$^{77}$Ge & $1/2^-$ & 0.160 & 0.193 & $\nu1\textrm{p}_{1/2}$ (94.7\%) \\
 & $9/2^+$ & 0.225 & 0.000 & $\nu0\textrm{g}_{9/2}$ (97.0\%) \\
\hline
 & $3/2^-$ & $0-0.215$ & 0.014 & $\pi1\textrm{p}_{3/2}$ (95.6\%) \\
$^{77}$As & $5/2^-$ & 0.246 & 0.000 & $\pi0\textrm{f}_{5/2}$ (92.5\%) \\
 & $9/2^+$ & 0.475 & 0.471 & $\pi0\textrm{g}_{9/2}$ (86.5\%) \\
\hline
 & $1/2^-$ & 0.000 & 0.227 & $\nu1\textrm{p}_{1/2}$ (93.6\%) \\
$^{77}$Se & $9/2^+$ & 0.175 & 0.000 & $\nu0\textrm{g}_{9/2}$ (97.6\%) \\
 & $3/2^-$ & 0.239 & 0.677 & $\nu1\textrm{p}_{3/2}$ (85.5\%) \\
 & $5/2^-$ & 0.250 & 0.506 & $\nu0\textrm{f}_{5/2}$ (93.0\%) \\
\hline
 & $3/2^-$ & 0.000 & 0.083 & $\pi1\textrm{p}_{3/2}$ (98.5\%) \\
$^{77}$Br & $9/2^+$ & 0.106 & 0.072 & $\pi0\textrm{g}_{9/2}$ (93.3\%) \\
 & $5/2^-$ & 0.162 & 0.000 & $\pi0\textrm{f}_{5/2}$ (95.3\%) \\
 & $1/2^-$ & 0.167 & 0.426 & $\pi1\textrm{p}_{1/2}$ (87.6\%) \\
\hline
\end{tabular}\end{center}\end{table}

The main point of the present Letter is to see how the neutron vacancies
of the pfg subspace, extracted from experimental data, affect the magnitude
of the $0\nu\beta\beta$ NME's. As a side line one can also try to relate the
measured vacancies to the single-particle energies of this subspace. The
WS potential is a global parametrization of the nuclear mean field 
and is based on data on nuclei close to the beta stability line. This
potential produces a smooth and gentle variation of the single-particle
energies as a function of the proton and neutron numbers. On the other hand,
the WS potential is just an approximate substitute for the
Hartree--Fock (HF) or Hartree--Fock--Bogoliubov (HFB) methods to calculate
the self-consistent mean field. The HF and HFB methods can produce results 
that are very different from those of the WS potential. Their results are also
very much dependent on the two-body interaction used in the calculations (for
recent articles on these features see \cite{KAN06,GUO07}).

In \cite{KAN06} a discussion of the $N=40$ and $Z=40$ mean-field gaps between
the $1\textrm{p}_{1/2}$ and $0\textrm{g}_{9/2}$ single-particle orbitals was
carried out by using the HF method with two different two-body interactions.
Let us first discuss the situation for protons. It
was found that on the proton side for one of the interactions, VMS
(see Fig.~15 of \cite{KAN06}), the gap between the $1\textrm{p}_{1/2}$ and 
$0\textrm{g}_{9/2}$ orbitals disappeares around $Z=34$ and the two orbitals
become inverted in energy. For the other interaction, JW 
(see Fig.~16 of \cite{KAN06}), the gap between these two orbitals
diminishes when going from $Z=26$ to $Z=48$ but never disappears. Thus the
VMS-computed mean field closely corresponds to our adjusted proton basis where
the gap between the $1\textrm{p}_{1/2}$ and $0\textrm{g}_{9/2}$ orbitals
disappears.

On the neutron side the VMS based HF calculation from $N=24$ to $N=48$ produces 
a clear gap between the $1\textrm{p}_{1/2}$ and $0\textrm{g}_{9/2}$ orbitals, 
as seen in Fig.~3 of \cite{KAN06}. However, in \cite{KAN06} it was found that
the effect of this gap disappeared when the $T=1,J=0$ pairing interaction was
added through the HFB method. If one would like to simulate
these vacancies in the present simple WS+BCS calculations the WS mean field
should be modified such that the gap at $N=40$ closes. 

At this point it has to be stressed,
however, that in the present work we do not have to resort to the details of
the underlying mean field and the associated neutron single-particle energies
since we have already the experimental vacancies
available. Some further insight in the complexity of the self-consistent mean-field
calculations for Ge isotopes is given in \cite{GUO07}. There several standard
interactions were used in the Gogny-HFB and Skyrme-HF+BCS frameworks and very
different results, from triaxial to axially symmetric shapes, were obtained with
different interactions for $^{76}$Ge. In fact, looking at the simple Nilsson
diagram indicates that a tiny oblate deformation would suffice to close the 
$N=40$ gap.

After settling the problem with the occupation amplitudes of the
single-particle states we are ready to
compute the $2\nu\beta\beta$ and $0\nu\beta\beta$ NME's. As usual, we consider
the two extreme values of the axial-vector coupling constant, namely the
bare value $g_{\rm A}^{\rm b}=1.25$ and the strongly quenched value $g_{\rm A}=1.00$.
When calculating the $0\nu\beta\beta$ half-lives it is convenient to remove the
$g_{\rm A}$ dependence from the phase-space factor by redefining the NME as
\begin{equation} \label{eq:meprime}
\textrm{NME}' = \textrm{NME}\left( \frac{g_{\rm A}}{g_{\rm A}^{\rm b}}\right)^2 \ .
\end{equation}
These redefined nuclear matrix elements are the ones that are listed in
Tables \ref{t:results} and \ref{t:others}.

In Table~\ref{t:results} we list the adopted $g_{\rm pp}$ values as
extracted by comparing the measured $2\nu\beta\beta$ half-lives with
the computed ones. By using these values of $g_{\rm pp}$ we have calculated
the $0\nu\beta\beta$ NME's of (\ref{eq:meprime}) and we summarize their values in 
Table~\ref{t:results}. In these calculations we have included the 
higher-order terms of nucleonic weak currents and the
nucleon's finite-size corrections in the way described in 
\cite{ROD06,SIM99}. We have accounted for the short-range correlations by the
Jastrow and UCOM (unitary correlation operator method) correlators, as
discussed, e.g. in \cite{KOR07a,KOR07b,KOR08}.

\begin{table}[htb]
\caption{Matrix elements of (\ref{eq:meprime}) computed in this work for different
values of $g_{\rm A}$ and $g_{\rm pp}$. For the short-range correlations both 
the Jastrow and UCOM prescriptions have been used.}
\label{t:results}
\begin{center}\begin{tabular}{cccccccc}
\hline
 & & \multicolumn{3}{c}{Jastrow} & \multicolumn{3}{c}{UCOM} \\
 $g_{\rm A}$ & $g_{\rm pp}$ & $(M_{\rm GT}^{(0\nu)})'$ & $(M_{\rm F}^{(0\nu)})'$ & 
$(M^{(0\nu)})'$ & $(M_{\rm GT}^{(0\nu)})'$ & $(M_{\rm F}^{(0\nu)}$)' & $(M^{(0\nu)})'$ \\
\hline
1.25 & 1.12 & 2.288 & $-0.772$ & 2.779 & 3.385 & $-1.143$ & 4.112 \\
1.00 & 1.10 & 1.700 & $-0.579$ & 2.279 & 2.413 & $-0.818$ & 3.231 \\
\hline
\end{tabular}\end{center}\end{table}

Our computed results of Table~\ref{t:results} can be compared with the results of
other recent works in the field. A selection of recent calculations including
the Jastrow and the UCOM correlator is given in Table~\ref{t:others}. The
second and third columns of this table give the results of \cite{KOR08} where
exactly the same methods as here were applied, the only difference being the
use of a different set of single-particle energies, where the neutron
0g$_{9/2}$ orbital was shifted a good one MeV to better reproduce the low-energy
spectra of $^{77}$Ge and $^{77}$Se in a BCS calculation. By comparing the
results of these two calculations in Tables~\ref{t:results} and \ref{t:others}
one notices a significant reduction in the value of the total $0\nu\beta\beta$
matrix element $M^{(0\nu)}$. Furthermore, the T\"ubingen results \cite{ROD07,SIM08} 
are consistent with the results of \cite{KOR08}.

\begin{table}[htb]
\caption{Values of the matrix element $(M^{(0\nu)})'$ of (\ref{eq:meprime}) 
obtained in some other recent works. Here (J) stands for Jastrow and (U) for UCOM.}
\label{t:others}
\begin{center}\begin{tabular}{ccccccc}
\hline
 $g_{\rm A}$ & (J)\cite{KOR08} & (U)\cite{KOR08} & 
(J)\cite{ROD07,SIM08} & (U)\cite{SIM08} & (J)\cite{MEN08} & (U)\cite{MEN08} \\
\hline
1.25 & 4.029 & 5.355 & 4.68 & 5.73 & 2.30 & 2.81 \\
1.00 & 3.249 & 4.195 & 3.33 & 3.92 & - & - \\
\hline
\end{tabular}\end{center}\end{table}

The results of the shell model \cite{MEN08} are the smallest in 
Table~\ref{t:others}. Interestingly, our present results for the Jastrow correlator,
$(M^{(0\nu)})'=2.779$, and for the UCOM correlator, $(M^{(0\nu)})'=4.112$,
are closer to the shell-model result than the previous values
quoted in \cite{KOR08}. The reduction of the magnitude of the pnQRPA 
calculated NME, which yields a value close to the shell-model result, is 
significant and deserves the further detailed study performed below.

\begin{figure}[htb]
\begin{center}
\includegraphics[width=12cm]{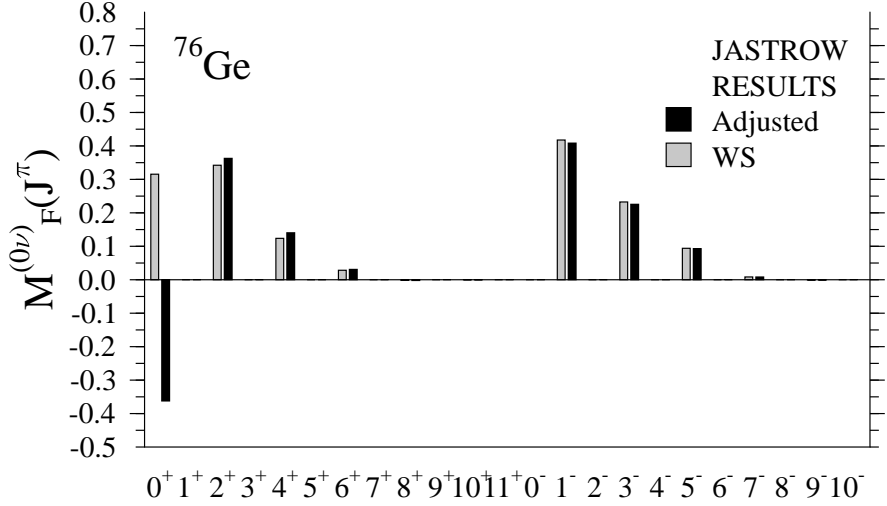}
\caption{Multipole decomposition of the Fermi matrix element with Jastrow
short-range correlations. The black bars correspond to the
modified neutron and proton occupations.}
\label{fig:MF}
  \end{center}
\end{figure}

\begin{figure}[htb]
\begin{center}
\includegraphics[width=12cm]{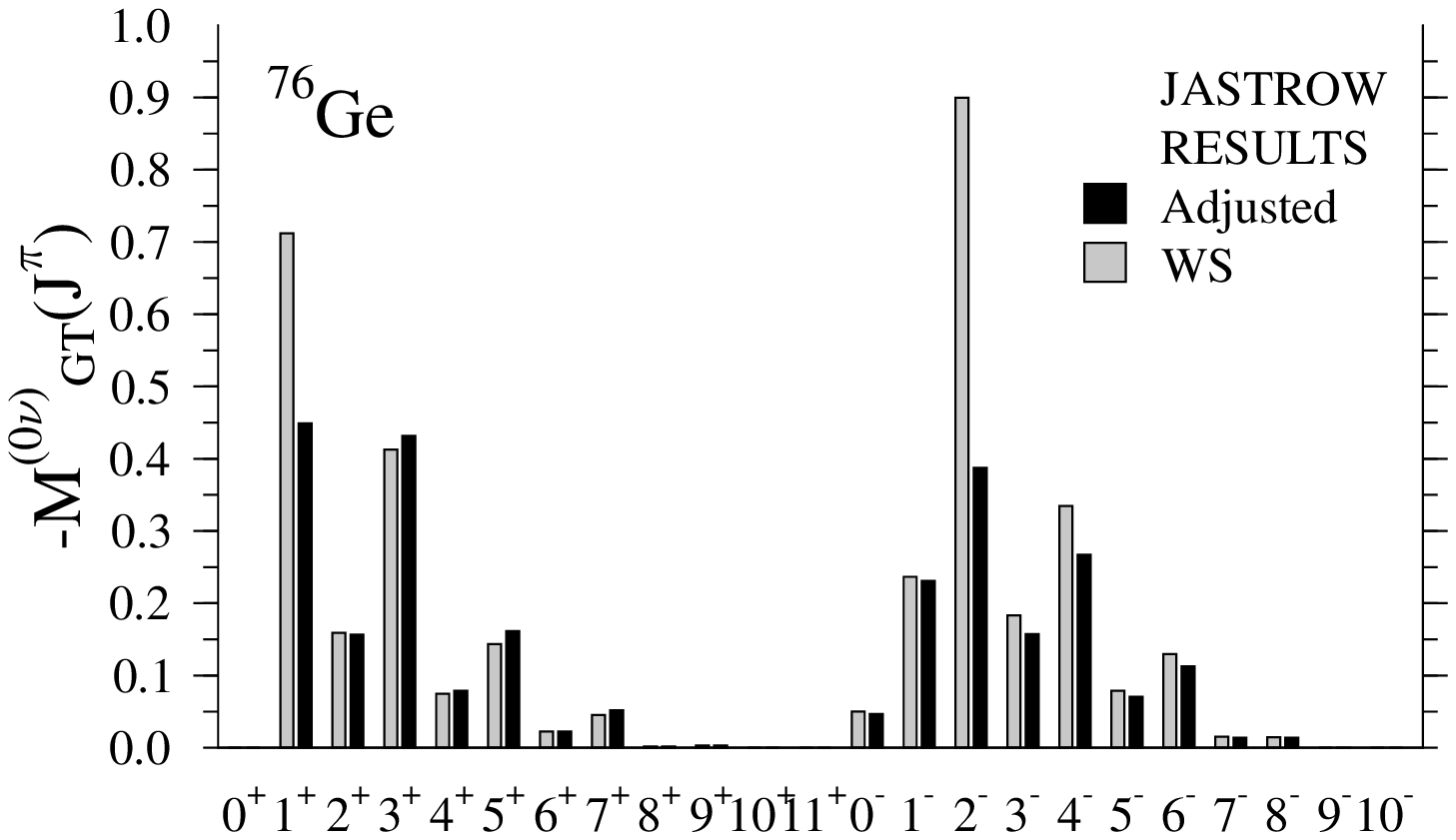}
\caption{The same as Fig.~\ref{fig:MF} for the Gamow--Teller matrix element.}
\label{fig:MGT}
  \end{center}
\end{figure}

The reason for the reduction of the magnitude of the $0\nu\beta\beta$ NME can be 
summarized by looking at the multipole decomposition of the NME. As an example we
use the Jastrow correlated NME's. For the
Fermi matrix element the reduction stems from the $0^+$ intermediate states,
as seen in Fig.~\ref{fig:MF}. From Fig.~\ref{fig:MGT} we see that for the 
Gamow--Teller matrix element $M_{\rm GT}^{(0\nu)}$ the significant changes concentrate
on the $1^+$ and $2^-$ contributions. In the present calculations the $1^+$ and $2^-$
contributions are $0.448(1^+)$ and $0.388(2^-)$ whereas in the previous work
\cite{KOR08} they read $0.712(1^+)$ and $0.899(2^-)$. Thus the $1^+$ contribution
has reduced by 37\% and the $2^-$ contribution by 56\%. In the old calculation
the contribution of the $2^-$ was dominant but now it has reduced below the $1^+$
contribution. In both calculations the $1^+$ multipole has several important 
contributions whereas the $2^-$ contribution is coming almost solely from the
first $2^-$ state. Hence the wave function of the $2^-_1$ state plays a key role
when seeking the reason for the reduction of the magnitude of the 
Gamow--Teller matrix element.

\begin{table}[htb]
\caption{Beta-minus decay $\log ft$ values for transitions from the $2^-_{\rm g.s.}$ 
of $^{76}$As to the ground state and one- and two-phonon states in $^{76}$Se.}
\label{t:logft}
\begin{center}\begin{tabular}{cccccc}
\hline
 & $0^+_{\rm g.s.}$ & $2^+_1$ & $0^+_2$ & $2^+_2$ & $4^+_1$ \\ 
\hline
Exp. & 9.7 & 8.1 & 10.3 & 8.2 & 11.1 \\
Present calc. & 9.7 & 7.4 & 9.0 & 8.4 & 10.7 \\
\cite{KOR08} & 9.0 & 7.7 & 9.2 & 8.7 & 10.9 \\
\hline
\end{tabular}\end{center}\end{table}

The quality of the lowest $2^-$ state in the intermediate nucleus $^{76}$As can be
tested by computing the $\beta^-$ decay $\log ft$ values for transitions from this
state to the ground state and one- and two-phonon states in $^{76}$Se. The obtained
results are compared with the data and the calculations of \cite{KOR08} in 
Table~\ref{t:logft}. As can be seen there is a drastic improvement in the 
$\log ft$ value of the ground-state-to-ground-state transition. This transition
tests exclusively the $2^-_1$ wave function whereas the rest of the transitions
depend also on the final-state wave function, built from the $2^+_1$ collective
phonon in the $^{76}$Se nucleus. It is worth pointing out that in the present 
calculation the quasiparticle spectrum is more compressed than in the calculation 
of Ref.~\cite{KOR08}. This increases the collectivity of the $2^+_1$ state in 
$^{76}$Se and thus results in smaller effective charges when trying to reproduce 
the data on the E2 transition probability from this state.

The single $\beta^-$ decay is a non-trivial way to check the reduction in the
$0\nu\beta\beta$ NME: The $\beta^-$ NME is reduced by 58\% from the old value
\cite{KOR08}. This is in nice agreement with the 56\% reduction in the $2^-$
contribution to the $0\nu\beta\beta$ NME. 
The main component of the wave function driving both 
transitions is the proton 0f$_{5/2}$ orbital coupled to the neutron 0g$_{9/2}$
orbital. In the present calculation the wave function of the $2^-_1$ state 
is more fragmented and thus reduces the pnQRPA amplitude responsible for 
the transitions. On the other
hand, the occupation of the neutron 0g$_{9/2}$ orbital has increased which also
reduces the decay amplitudes since they are proportional to the
emptiness of $\nu0\textrm{g}_{9/2}$. Similar considerations, though in a 
more complicated way, apply to the intermediate $1^+$ contribution.

Our calculations show that the main contributions to the $0\nu\beta\beta$ NME
come from inside the pfg subspace. The implementation of the experimental 
occupations in the pnQRPA calculation brings the pnQRPA results closer to 
the shell-model results of \cite{MEN08}. 
The small contributions from outside the pfg subspace partly explain the
deviations from the shell model result. In \cite{CAU08b} the effect of expanding the
shell-model single-particle basis was examined. Using the pfg subspace plus
two-particle--two-hole excitations from the 0f$_{7/2}$ orbital it was concluded
that these 2p-2h excitations increase the magnitude of $M^{(0\nu)}$ by at most
20\%. It still remains an open question how the differences between the shell-model 
and pnQRPA matrix elements tie to the omitted single-particle orbitals 
in the shell model and the shell-model occupations of the pfg subspace.

Finally, our presently computed variations
$(M^{(0\nu)})'=2.279-2.779$ (Jastrow) and $(M^{(0\nu)})'=3.231-4.112$ (UCOM) in the
$0\nu\beta\beta$ NME can be converted to the following half-life limits
\begin{eqnarray}
\label{eq:finalme}
t_{1/2}^{(0\nu)} &=& (5.36-8.04)\times 10^{24}\,\textrm{yr}/
(\langle m_{\nu}\rangle[\textrm{eV}])^2 \ (\textrm{Jastrow}) \ , \\
t_{1/2}^{(0\nu)} &=& (2.45-4.00)\times 10^{24}\,\textrm{yr}/
(\langle m_{\nu}\rangle[\textrm{eV}])^2 \ (\textrm{UCOM}) \ .
\end{eqnarray}

In this Letter we have performed a pnQRPA calculation of the nuclear matrix
elements involved in the neutrinoless double beta decay of $^{76}$Ge. We have
used a microscopic two-nucleon interaction in a realistic model space,
and the calculations exploit the occupation amplitudes extracted from the recently
available data on the neutron vacancies in $^{76}$Ge and $^{76}$Se. The
subsequently calculated $0\nu\beta\beta$ nuclear matrix elements are smaller
in magnitude than the ones obtained in a standard calculation using the
Woods--Saxon based single-particle occupations. This stems from the reduction
in the contributions of the $0^+$, $1^+$ and $2^-$ intermediate states, with a
special emphasis on the first $2^-$ state. These changes are
related both to the revised occupations and to the changes in the pnQRPA 
amplitudes that derive from the revised occupations.

{\bf Acknowledgements:} This work has been partially supported by the 
National Research Council (CONICET) of Argentina and by the
Academy of Finland under the Finnish Centre of Excellence Programme 
2006-2011 (Nuclear and Accelerator Based Programme at JYFL). We thank
also the EU ILIAS project under the contract RII3-CT-2004-506222.
One of the authors (O.C.) gratefully thanks for the warm hospitality 
extended to him at the Department of Physics of the University of 
Jyv\"askyl\"a, Finland. We acknowledge the very fruitful discussions with Prof. 
J.P. Schiffer. We also thank Dr. M. Kortelainen  and Mr. M.T. Mustonen for 
making their computer codes available to us.

\end{document}